\documentstyle[12pt]{article}
\begin{document}
\baselineskip=20pt
\def\s{\section}
\def\ss{\subsection}
\def\sss{\subsection}
\def\ni{\noindent}
\def\hf{\hfill\break}
\def\i{\item}
\def\be{begin{equation}}
\def\ee{end equation}
\def\ni{\noindent}
\noindent
\vspace*{0.8cm}

\vspace*{2.5cm}

\begin{center}
{\large\bf A 3 - dimensionally modulated}

{\large\bf structure in a chiral}

{\large\bf smectic-C liquid crystal}
\vspace*{0.2cm}

{\bf P.A. Pramod, R. Pratibha and N.V. Madhusudana}

\vspace*{0.2cm}
Raman Research Institute, C.V. Raman Avenue, Bangalore 560 080, India
\end{center}
\vspace*{0.8cm}

\noindent
{\it Key words}

\noindent
Liquid Crystal, Twist Grain Boundary Phase (TGB), Smectic C$^*$.
\vspace*{0.4cm}

\noindent
{\bf Abstract}:

In this article we report the discovery of a new Twist Grain
Boundary phase. This phase is characterised by a 2-dimensional
undulation of the smectic C$^*$  like blocks in the form of a
square lattice. We suggest that this three dimensionally
modulated structure, which was not anticipated by theory, owes its
origin to chiral interactions.

\newpage
\noindent
{\bf Introduction}:

The formal analogy between superconductors and smectic
liquid crystals was invoked by de Gennes$^1$  to predict the
possibility of an intermediate phase with a lattice of
dislocations in smectics.  Goodby {\it et al}$^2$ discovered
such a ~structure ~in a ~highly ~chiral ~liquid ~crystal. ~This
"Twist ~Grain ~Boundary" ~(TGB$_A$) phase consists of a helical ~stack ~of
~blocks ~of ~smectic A (S$_A$) liquid crystals, separated by
grain boundaries made of an array of screw dislocations (Fig.1a), in
accordance with a  structure which had been worked out by Renn and
Lubensky$^3$.  Unlike superconductors, smectic liquid
crystals can have other modifications like the smectic C (S$_C$) in
which the molecules are tilted with respect to the layer normal
and the smectic C$^*$ (S$_{C^*}$) in which  the tilt direction
has a helical arrangement about the layer normal.  Although TGB phases with S$_C$
like blocks (TGB$_C$) (Fig.1b) and S$_{C^*}$ like blocks (TGB$_{C^*}$) have been
theoretically predicted$^{4,5}$, only the TGB$_C$ phase has been
experimentally characterised in some detail$^{6,7}$.  Liquid
crystals are rather soft, and can exhibit novel geometrical
structures. We have found in a binary mixture a new TGB phase  which has a 2D undulation of the
S$_{C^*}$ blocks in the form of a square lattice. 

\noindent
{\bf Experimental}:

The new phase was found in binary mixtures of the chiral
compound 4-(2'-methyl butyl phenyl 4'-n-octyl
biphenyl-4-carboxylate (CE8) and
2-cyano-4-heptylphenyl-4'-pentyl-4-biphenyl carboxylate (7(CN)5)
which have very similar lengths and molecular structures.  On
heating, CE8 exhibits the phase sequence (with temperature in $^o$C) : crystal 67 S$_{I^*}$
70 S$_{C^*}$ 85 S$_A$ 134.6 N$^*$ 140.5 I where N$^*$
stands for chiral nematic and I for isotropic phases. On the
other hand, 7(CN)5 has a wide nematic  range  (crystal 45
N 102 ~I) and Xray studies have shown that it has a strong
skew cybotactic (S$_C$ like) short range order$^8$. The phase
diagram of the binary mixtures is shown in Fig 2.  The TGB
phases are found only in mixtures with $\sim$ 5 wt \% to 45 wt
\% of 7(CN)5. Most of the physical studies were conducted on a
mixture with about 36 wt \% of 7(CN)5, which exhibits the
following (known) phases on cooling: I 121.7  N$^*$ 76.8 TGB$_A$. Observations using a polarising
microscope show that as the sample is cooled further to 63$^o$C,
there is a distinct transition from TGB$_A$ phase to another phase, in which large
patches develop a square grid pattern. As the temperature is
lowered below 59$^o$C the grid becomes less distinct.

In a cell whose glass plates are pretreated with  polyamide and
unidirectionally rubbed so that the nematic director has a planar
alignment, the TGB$_A$ phase exhibits a  Grandjean plane texture$^1$
similar to the cholesteric phase. In such a sample the boundary condition
ensures that the helical axis is perpendicular to the glass plates. The local
changes in the cell thickness  produces sympathetic  variations in
the helical pitch producing the Grandjean plane texture. As the director
alignment is fixed on the glass surfaces, only an integral number of
half pitches can be accomodated between the plates. The number changes by 
unity across each Grandjean Cano (GC) dislocation line. As the temperature
is lowered the texture goes over to a well aligned square grid pattern, 
one of the axis of which is parallel to the rubbing direction. In wedge
shaped samples the GC lines  are seen in both the TGB$_A$ (Fig.3a) and the new
phase with the square grid, demonstrating that there is a helical twist
normal to the plates in both the cases (Fig.3b).  As the
temperature is lowered to 59$^o$C, the GC lines become
highly distorted and appear to get anchored at surface irregularities. 
The spacing between GC lines
increases only down to 59$^o$C below which the
irregular lines are not affected by temperature. 
The  square  grid is a pseudomorphotic (or metastable) texture below
59$^o$C and is erased by  a displacement of the cover slip,
to produce a texture
characteristic of the  S$_{C^*}$ phase. As the temperature is
increased, the square grid texture is recovered at 59$^o$C, as
are the  distinct and straight GC lines in the wedge shaped
sample. These observations show that there is a new  phase which occurs between
the TGB$_A$ and S$_{C^*}$ phases. It is characterised by a
helical arrangement whose axis is  normal to the glass plates
for planar alignment and a square grid modulation in
the orthogonal plane. It is a thermodynamically distinct phase
which appears both on cooling from the TGB$_A$ phase and  on
heating from the S$_{C^*}$  phase.

We have also prepared cells whose glass plates are coated with a thin layer of glycerine
which produces a degenerate planar anchoring. Again the square grid
texture appears in both the heating and cooling runs. Of course in a wedge
shaped sample no GC lines form due to the degeneracy in the boundary 
condition, and the lattice spacing of the square grid is found to be
independent of the thickness of the sample. Thus, surface
anchoring is not important in stabilising the modulated
structure. In order to characterise the structure in greater
detail, we have conducted the following experiments:

\begin{description}
\item (a) ~~The diffraction pattern of a laser beam  from the
structure brings out  the
underlying square grid distortion which results in a
periodic variation of the effective refractive index (Fig 3c).

\item (b) ~~In cells prepared with homeotropic boundary
conditions in which the director preferentially   aligns normal to
the glass plates, the TGB$_A$ phase is characterised by a
filamentary texture$^9$, each filament corresponding to a
rotation of the director by  $\pi$ radians, 
and the width of the filament $\approx$ p/2 where p is the pitch of the TGB
helix$^9$. When the sample is cooled to 63$^o$C, the filaments very clearly
develop an undulatory structure  (Fig 3d) with a periodicity which
roughly corresponds to that of the square grid observed in the
planar geometry. As the temperature is lowered to 59$^o$C, i.e, in the
S$_{C^*}$ phase, the filaments disappear, but on reheating they
reappear with the undulatory  structure. The structure straightens out when
the temperature is raised to 63$^o$C.

\item (c) ~~When small drops of the liquid crystal are deposited
on a glass plate  treated for homeotropic alignment, in the 
S$_{C^*}$  phase concentric GC rings are seen.
As the temperature is raised to 59$^o$C, undulatory filaments
grow as arcs with their centres at the geometric centre of the
drop.

\item (d) ~~Xray scattering studies on  25$\mu$m thick aligned samples
taken between  etched cover slips show that in the TGB$_A$ phase
there is a relatively  (compared to N$^*$) sharp ring corresponding to the S$_A$ layer
spacing$^2$. As the temperature  is lowered below 63$^o$C the ring
expands showing that the molecules are now tilted 
in the layers. Below 59$^o$C,  in addition to
the ring, four relatively strong spots  are also seen. 
Probably they indicate an unwound region of S$_{C^*}$ liquid
crystal near the glass surfaces. In the present experiment, we
are unable to verify if the TGB structure in the new phase is
commensurate  or otherwise.

\item (e) ~~The effect of an external AC electric field was
studied in different geometries. Under the action of a~10KHz field 
applied along the TGB helical axis in the planar geometry, the
dark regions separating the bright ones (see Fig.3b) become very
thin and straight and remain intact even at 30V/$\mu$m. In the
homeotropic geometry, under an appropriate setting of crossed
polarisers, the undulatory filaments are seen to have periodic
dark and bright bands along the length. If the filament is roughly parallel to a
pair of wires between which a transverse electric field is
applied, the dark bands expand and the bright ones shrink.   
When a very low frequency
(1Hz) square wave voltage is applied between ITO coated
plates in the same geometry, for a field $\sim$ 5V/$\mu$m, 
the filaments become broader and straight. Further, a narrow dark band in the
{\it centre} of the filament shows a spatially periodic intensity
modulation  along the length which responds at the frequency of the applied voltage.
This is  best seen when the polariser is set nearly orthogonal
to the `axis' of the filament. On the other hand, a similar voltage
applied to a TGB$_A$ filament shows a continuous dark band in
the centre. 
\end{description}

\noindent
{\bf Discussion}:

Based on the above observations, we propose that the
intermediate phase is TGB$_{C^*}$ in nature, i.e., there is a
helical arrangement of tilted molecules within each  S$_{C^*}$ like
block. In addition, these TGB$_{C^*}$ blocks have a two
dimensionally undulating structure such that it forms a
square grid. A schematic diagram of such a structure
is shown in Fig. 4.

The two dimensional modulation with a square lattice is a
consequence of the uniaxial nature of the TGB phase which has a
helical twist. The electric field experiments
on the material with negative dielectric anisotropy can be understood if there is a helical twist {\it within each block},
the field induced unwinding of which  produces
solitons$^1$ which appear like thin lines  of the square grid
or  a periodic intensity modulation along the length of the straightened
filaments. Indeed the observed undulatory nature of the filament
in the absence of field is a `side view' of the 2-D undulating
structure of the medium. The proposed structure is rather 
non-uniform with helical axes characteristic of a TGB smectic as
well as in the orthogonal plane, ie., in the blocks. This is
reminiscent of the blue phases exhibited by short pitched
cholesterics close to the transition point to the isotropic
phase$^1$. However, the present structure is anisotropic and
does not have cubic symmetry. As the smectic layer normals of the
blocks rotate across the grain boundaries, the structure is
highly non-uniform. The two dimensional undulation was not
anticipated  in the theoretical models of the TGB$_{C^*}$
phase$^5$. Since the grain boundaries also  undulate along
with the entire structure, we call the new phase the undulated
TGB$_{C^*}$ (UTGB$_{C^*}$) phase.

The temperature variations of the TGB pitch (measured using
the spacing between GC lines  in a wedge shaped sample) and the lattice
spacing of the square grid (measured using the optical diffraction
pattern) are shown in Fig 5 and Fig 6 respectively. 
The TGB pitch increases as the 
temperature is lowered, the rate of variation becoming very large in the
UTGB$_{C^*}$ phase. Measurements on the TGB$_C$ phase also
show a similar trend$^6$. On the other hand, the lattice
spacing of the square grid decreases quite sharply as the
temperature is lowered from TGB$_A$ to UTGB$_{C^*}$ transition point,
and levels off at lower temperatures. Indeed the pitch in
the  S$_{C^*}$ phase roughly corresponds to
the lattice spacing at the lowest temperatures of UTGB$_{C^*}$ phase.

The physical origin of the UTGB$_{C^*}$ phase with its highly non-uniform structure 
is obviously of interest. It would appear that the
elastic energy cost of the deformations involved  would normally make
such a structure unlikely. However, we must remember that the
TGB$_A$ phase itself has a considerable   non-uniformity, with almost
perfect S$_A$ blocks separated  by  highly defected grain
boundaries with  screw dislocations. These two  parts of the structure are
so different that an anisotropic interfacial energy may be
invoked between the blocks. The tilting of the
molecules at the transition point may be expected to produce a helical twist
along the smectic  layer normal in a block$^1$. It is easy to see
that the angle made by the director  with the grain boundary varies
along the layer normal if the grain boundary  remains flat. This would
cost extra energy which can be thought to arise from the fact that the director
distortion {\it across the grain boundary} is no longer a pure twist.
This energy can be lowered if the grain boundary and with it the
blocks undulate along the
smectic layer normal (Fig.4).  This ensures that the director is parallel to the
grain boundary, reducing  the grain boundary energy, and of course the
helical twist in the block  is favoured by the gain in the chiral
energy. Note that the dislocations are no longer pure screw dislocations$^{10}$.  The undulation instability takes place along two
mutually orthogonal directions in view of the uniaxial symmetry
of the TGB structure. This means that in most of the blocks the
smectic layer normal  makes non zero angles with the two axes
of the square grid. Indeed it is possible to show that for an
appropriate set of parameters, i.e, the anisotropic grain boundary energy, 
the chiral term and the other
elastic constants, the undulatory structure is energetically more
favourable than the TGB$_C$ structure in which  there is neither a
helical twist nor undulations in the blocks.

Thus the UTGB$_{C^*}$ phase is characterised by helical axes
both along and normal to the S$_{C^*}$ layers and it naturally
occurs between the TGB$_A$ in which the helical axis is parallel
to the S$_A$ planes, and S$_{C^*}$ in which it is normal to the
S$_{C^*}$ planes. Very recently we have found that the UTGB$_{C^*}$  phase occurs in a
couple of other systems. Detailed experimental and theoretical
results on this phase, which is perhaps one of the most non-uniform liquid crystalline
phases found as yet, will be published elsewhere.

\noindent
{\bf Acknowledgments}:

Our thanks are due to Yashodhan Hatwalne for numerous discussions
on the TGB phases.

\vspace*{0.4cm}

\noindent
Correspondence and requests for materials to N.V.M
(e-mail:nvmadhu@rri.ernet.in).

\newpage
\noindent
{\bf Figure Legends}\\
\begin{enumerate}

\item[{Fig.1}] (a) Schematic representation of the TGB$_A$ structure. The 
direction of the average orientation of the molecules (director) is along
the local layer normal. (b) Schematic representation of the TGB$_C$
structure (after reference 7).
{\bf N} is the layer normal and {\bf n} is the director. {\boldmath $\tau$} is the helical
axis.

\item[{Fig.2}]Phase diagram of the binary mixtures of CE8 and
7(CN)5. Note that the TGB phases occur only in the mixtures and
the new phase (UTGB$_{C^*}$) occurs in a narrow range of
compositions.

\item[{Fig.3a}]Photograph of the Grandjean Cano lines seen in the TGB$_A$
phase in a wedge shaped sample with pretreated glass plates as explained
in the text. 

\item[{Fig.3b}]Photograph of the square grid pattern (grid spacing $\sim$ 2.5 $\mu$m) seen in the
new phase at 60$^o$C in a wedge shaped sample. Note the two vertical
Grandjean-Cano lines
which are characteristic of a helical arrangement normal to the
plane of the figure.

\item[{Fig.3c}]Photograph of the diffraction pattern of a He-Ne laser beam produced by
the square lattice of the type shown in Fig.3b 

\item[{Fig.3d}]Photograph of the filamentary texture  just below
the transition to the UTGB$_{C^*}$ phase exhibited in samples whose
walls are pretreated for homeotropic alignment.  Note the
undulations in the filaments. (crossed polarisers, periodicity of
the undulation $\sim$ 4 $\mu$m).

\item[{Fig.4}]Schematic  diagram of the geometrical arrangement
of two neighbouring smectic C$^*$ blocks in the proposed
structure in the new phase. Note that the 2D-undulations in both
the blocks have the same orientations of the wavevectors. The
orientation of the smectic layer normal (large arrows) is
different in the two blocks, which are separated by a grain
boundary (not shown explicitly). The dotted areas representing smectic layers
have undulations only in the vertical plane. The helicity of
the director in the lower smectic C$^*$ block is shown by that
of the `nails'.

\item[{Fig.5}]Temperature variation of the pitch in the three
helical phases, viz., N$^*$, TGB$_A$ and UTGB$_{C^*}$. 

\item[{Fig.6}]Temperature variation of the lattice spacing
in the square grid phase.
\end{enumerate}

\newpage
\baselineskip=19pt
\begin{thebibliography}{99}
\bibitem{}De Gennes, P.G. \& Prost, J. The Physics of Liquid
Crystals (Clarendon Press, Oxford 1995)

\bibitem{}Goodby, J.W., Waugh, M.A., Stein, S.M., Chin, E.,
Pindak, R. \& Patel, J.S. Characterisation of a new helical
smectic liquid crystal. {\em Nature} {\bf 337}, 449-452 (1994).

\bibitem{}Renn, S.R. \& Lubensky, T.C. Abrikosov dislocation
lattice in a model of the cholesteric-to-smectic A transition.
{\em Phys.Rev} {\bf A 38}, 2132-2147 (1988).

\bibitem{}Renn, S.R. \& Lubensky, T.C. Existence of a sm-C grain
boundary phase at the chiral NAC point. {\em
Mol.Cryst.Liq.Cryst} {\bf 209}, 349-355 (1991).

\bibitem{}Renn, S.R. Multicritical  behaviour of Abrikosov
vortex lattices near the cholesteric -smectic A-smectic C$^*$
point.   {\em Phys.Rev A} {\bf 45} (2), 953-973 (1992).

\bibitem{}Isaert, N., Navailles, L., Barois, P. \& Nguyen, H.T.
Optical evidence of the layered array of grain boundaries in
TGB$_A$ and TGB$_C$ mesophases.  {\em J.Phys.II France} {\bf 4}
1501-1518 (1994).

\bibitem{}Navailles, L., Pindak, R., Barois, P. \& Nguyen, H.T.
Structural study  of the smectic C  twist grain boundary phase
(TGB$_C$) {\em Phys. Rev.Lett} {\bf 74}, 5224-5227 (1995).

\bibitem{}Madhusudana, N.V., Moodithaya, K.P.L. \& Suresh, K.A.
Effect of skew cybotactic structure on the optical properties of
a nematogen with a lateral cyano substituent. {\em Mol.Cryst}.
{\bf 99}, 239-247 (1983).

\bibitem{}Goodby, J.W., Waugh, M.A., Stein S.M., Chin, E.,
Pindak, R. \& Patel, J.S. A new molecular ordering in helical
liquid crystals. {\em J.Am.Chem.Soc III}, 8119-8125 (1989)

\bibitem{}Chaikin, P.M. \& Lubensky, T.C.,  Principles of
condensed matter physics, 539 (Cambridge University Press)
\end{thebibliography}
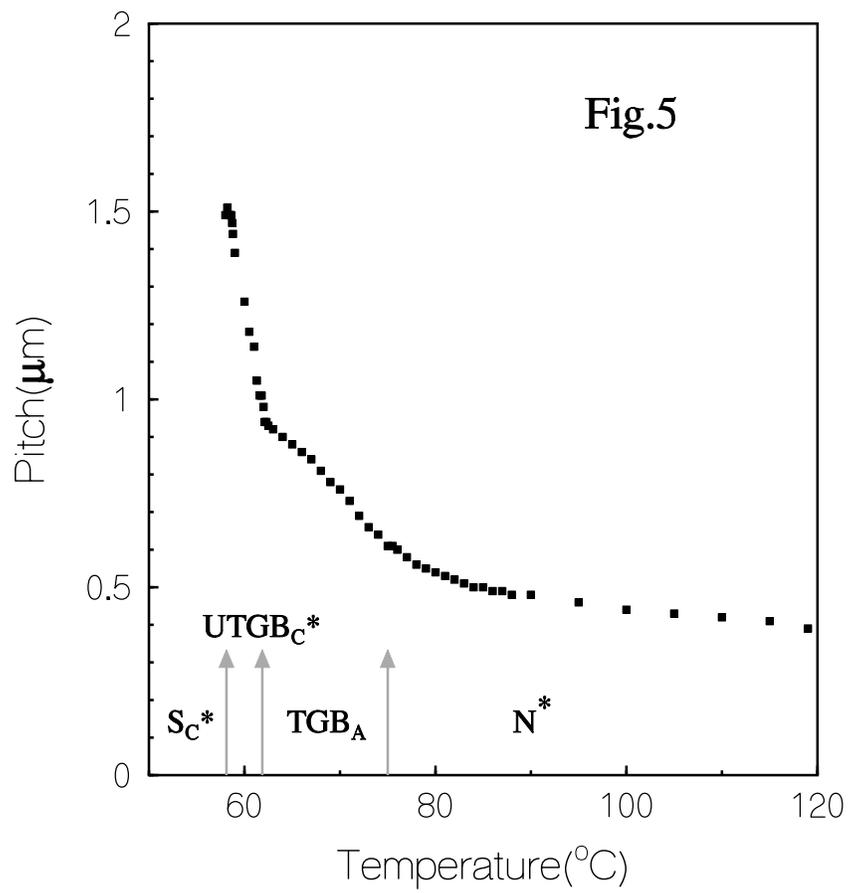
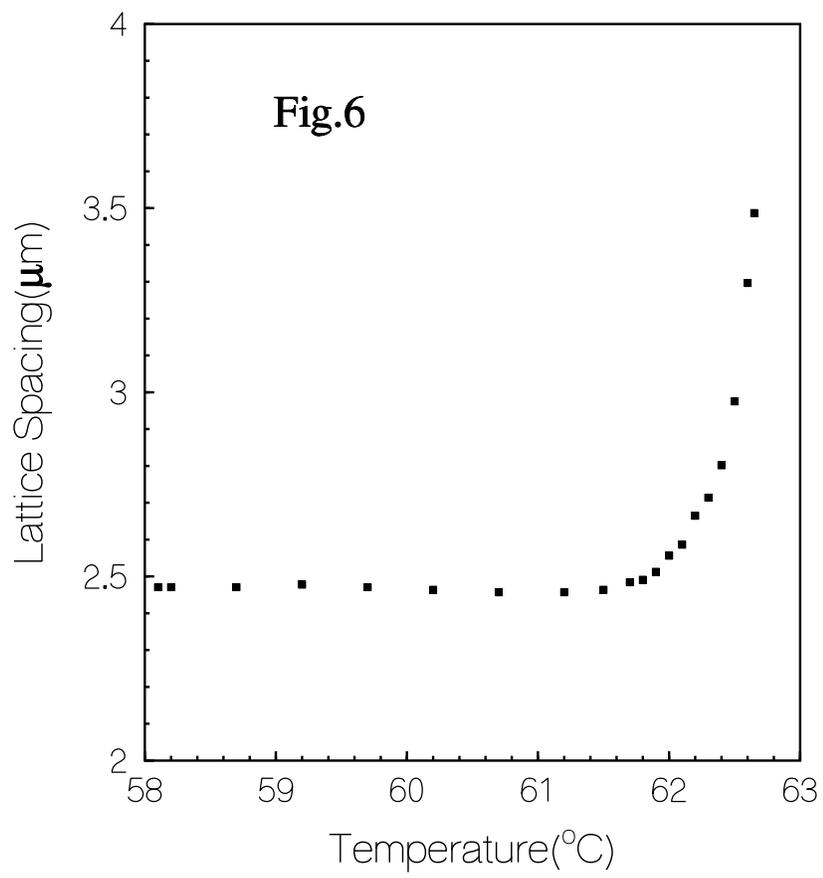
\end{document}